\documentclass[twocolumn]{aastex631}
\usepackage{amsmath, amssymb, xcolor}
\usepackage{hyperref}
\usepackage{float}

\begin{document}
\title{From Origins to Observables: Distinguishing Dark Compact Objects with Population-Level Microlensing Signatures}

\author{Joel Cortez Osuna}
\affiliation{Institute for Gravitation and the Cosmos, The Pennsylvania State University, University Park, PA 16802, USA}
\affiliation{Department of Physics, The Pennsylvania State University, University Park, PA 16802, USA}

\author{Sarah Shandera}
\affiliation{Institute for Gravitation and the Cosmos, The Pennsylvania State University, University Park, PA 16802, USA}
\affiliation{Department of Physics, The Pennsylvania State University, University Park, PA 16802, USA}

\begin{abstract}
While primordial black holes (PBHs) have long been a benchmark target for microlensing searches, the modern landscape of dark matter models suggests other, distinct, formation channels for compact objects made of dark matter. In the large class of self-interacting, dissipative models, dark matter has cooling channels that can enable fragmentation and gravitational collapse of some dark matter into compact objects, including black holes. The resulting populations have mass distributions, bias parameters, and abundance, spatial profile and velocity dispersion within the Milky Way that all differ from those of PBHs. We demonstrate that these population-level differences can leave imprints in the space of microlensing observables, with the differences in how the populations trace the dark matter giving the primary distinguishing lever. We discuss the possible overlap of microlensing signals from dark and baryonic lenses, and the complementarity of microlensing detection or constraints with other gravitational probes of novel populations of dark matter origin. 
\end{abstract}

\section{Introduction}
Gravitational microlensing, the transient amplification of a stellar source due to the deflection of light from the gravitational field of a massive object \citep{mao2008pos}, provides a unique observational probe of the isolated compact object population. Since microlensing depends on the gravitational field of the lens mass through the geometry and kinematics of the source-lens system rather than its luminosity, this technique can find faint or non-luminous lenses that would otherwise elude detection. As a result, microlensing has historically been the primary method to find isolated compact objects that may constitute a fraction of the dark matter, $f_{\text{DM}}$ \citep{paczynski1986gravitational, griest1991galactic, green:2026}. 

Early surveys such as the MACHO Project~\citep{alcock2000macho}, EROS~\citep{eros1997microlensing}, and OGLE~\citep{mroz2024microlensing}, commissioned in the 1990s and 2000s, pioneered the search for massive compact objects in the Galactic halo \citep{ Roulet_1997}. Collectively, these observations placed upper limits on the fraction of dark matter in compact halo objects across a wide range of lens masses. Microlensing surveys toward the Magellanic Clouds constrained the mass range of $10^{-7} - 10^3$ $M_{\odot}$, placing upper bounds on $f_{\text{DM}} \lesssim 1 - 10\%$ \citep{alcock2001macho}, while later observations towards the Galactic bulge provided additional constraints in the mass range $10^{-6} - 10^{-3}$  $M_{\odot}$, again finding a limit of $f_{\text{DM}} \lesssim 1 - 10\%$ \citep{niikura2019constraints,Mroz:2025}. The constraining power of these data was limited by survey cadence, line-of-sight, and model assumptions.

With upcoming large-scale time-domain surveys such as Rubin's Legacy Survey of Space and Time (LSST) \citep{sajadian2019predictions, abrams2025microlensing} and the Roman Space Telescope Galactic Bulge Time Domain Survey (GBTDS) \citep{penny2019predictions}, microlensing constraints will improve dramatically. These next-generation surveys will have improved sensitivity, enabling a higher detection rate of lenses across a wide range of masses, from exoplanets to stellar remnants like black holes. With a projected 10-year survey towards the Galactic Center and wide-field monitoring of over billions of stars \citep{street2018unique}, LSST is expected to detect $10^{4}$ microlensing events per year. The Roman Space Telescope's projected 5-year survey and high-frequency cadence observations are similarly expected to provide nearly $10^4$ microlensing events over its bulge survey \citep{johnson2020predictions}. The increase in variety and number of detections will shift microlensing to a population-level inference probe. As a result, constraints on the fraction of dark matter in compact objects are forecast to improve by as much as several orders of magnitude, depending on the mass range \citep{drlica2019probing, winch2022using, pruett2022primordial,fardeen2024astrometric,Romao:2025}.  

Primordial black holes (PBHs), theorized to have formed from early-universe overdensities \citep{zel1967hypothesis, hawking1971gravitationally, carr1974black, carr1975primordial}, have historically dominated the conversation on low-mass black holes in microlensing and gravitational wave searches \citep{carr2024history, bagui2025primordial} and have been included in microlensing simulation frameworks and forecasts \citep{pruett2022primordial, derocco2024revealing, fardeen2024astrometric}. Whether PBHs are taken to be all of the dark matter or just a fraction, they are generally assumed to trace the overall dark matter distribution. Their spatial and velocity distributions within the Milky Way are inherited from the dark matter profile of the galaxy. In most scenarios, the mass distribution of PBHs is primarily determined by the scale on which the primordial power spectrum is assumed to have a sufficiently large amplitude for the black holes to form by direct collapse of over-dense regions. Since there are no other direct probes of the primordial spectrum on such small scales, the only constraints on what spectra can be considered come from restrictions on inflation proceeding \citep{Cole:2023wyx}. The PBH abundance is a function of both the amplitude assumed for the primordial perturbations and the equation of state of the hot big bang universe at the time the wavelengths of the overdensities are of order the Hubble scale.  

Current microlensing simulations and forecasts largely assume that any non-baryonic compact object population can be represented using PBH models. However, a large class of particle dark matter models also predict populations of dark compact objects, with distinct formation mechanisms that generate population demographics fundamentally distinct from PBH models. Given that microlensing observables depend directly on population properties, the unique formation channels of different dark compact object (DCO) populations may produce statistically distinguishable signatures in the microlensing observables parameter space.

Beyond the PBH scenario, dark matter that consists of more than one interacting particle may have cooling channels that allow the formation of compact objects \citep{Mohapatra:1999ih}. There are motivations to consider this broad class of models both from particle physics \citep{chacko2006natural, martin2010supersymmetry, csaki2016tasi, chacko2017cosmology, craig2017cosmological,Lankester--Broche:2025rkm} and astrophysical observations, especially the diversity of rotation curves of galaxies and other small-scale structure observations \citep{adhikari2022astrophysicaltestsdarkmatter,kamada2017self, ren2019reconciling}. Compact object formation in dissipative dark matter scenarios shares features of star formation in baryonic matter. Typically, only a small fraction of the dark matter will end up in compact objects, which will be concentrated in regions where the overall matter density met the chemical criteria for sufficient cooling \citep{buckley2018collapsed}. Although those regions will not be identical to those where Standard Model cooling is effective, dark compact objects formed from dissipative dark matter will therefore more closely trace baryonic compact objects than the full dark matter distribution. The abundance and type of objects formed when some dark matter can cool depends on details of the dark sector chemistry. Several different scenarios with different cooling channels and associated spectra of compact objects have been studied, demonstrating that dark white dwarfs \citep{Ryan:2022hku}, dark neutron stars \citep{hippert2022mirror}, and dark black holes (DBHs) spanning a wide range of masses \citep{DAmico:2017lqj,shandera2018gravitational,Chang2019,Gurian:2022nbx,Bramante:2023ddr} are all possible. Since microlensing is only sensitive to the mass, spatial, and kinematic distribution of dark compact objects, in this paper we will refer to this whole class as DBHs for simplicity.

\begin{figure}[t]
    \centering
    \includegraphics[width=0.85\linewidth]{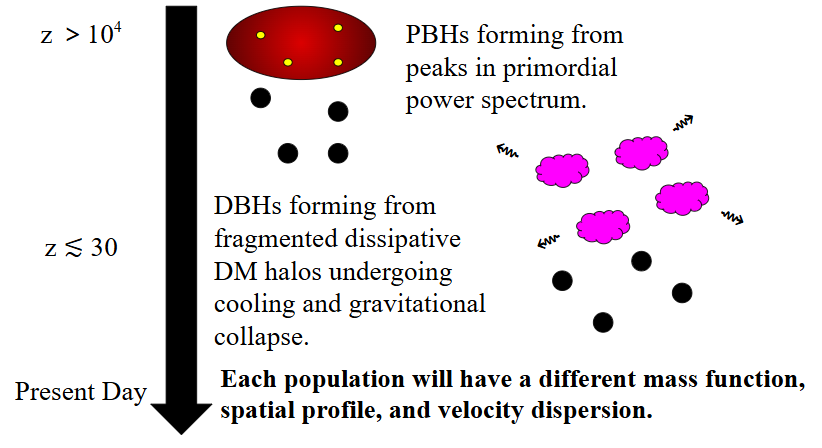}
    \caption{Primordial black holes (PBHs) and dark black holes (DBHs) have different formation channels, resulting in populations that differ in mass, abundance, and distribution. The PBH mass function is determined by very early universe cosmology. PBHs behave as collisionless dark matter, inheriting their spatial and kinematic distributions from the host dark matter halo. By contrast, the DBH mass function and abundance is determined by the dark sector cooling efficiency, and the spatial and kinematic distributions may reflect the formation or merger history of the host galaxy. The differences in formation processes leave signatures on microlensing observables. \label{fig:FormationSchematic}  }
\end{figure}

The qualitative differences between PBH and DBH formation are illustrated in Figure \ref{fig:FormationSchematic}.

\subsection{Microlensing Observables}
Gravitational microlensing events occur when a compact object of mass $M$ at distance $D_{\text{L}}$ from the observer passes through the line of sight of a stellar source at distance $D_{S}$, bending its light and producing multiple images that cannot be resolved by current telescopes. As a result, the microlensing signature is a time-symmetric amplification of the magnitude of a stellar source \citep{mao2008pos}. The timescale of these events is characterized by the Einstein crossing time, 
\begin{equation}
    t_{E}  = \frac{\theta_{E}D_{L}}{v_{\perp}},
\end{equation}
where $v_{\perp}$ is the relative transverse velocity between the lens and the source, and
\begin{equation}
    \theta_{E} = \sqrt{\frac{4GM}{c^{2}}\left(\frac{1}{D_{L}} - \frac{1}{D_{S}}\right)}
\end{equation} 
is the angular Einstein radius \citep{einstein1936lens}, where $c$ is the speed of light and $G$ is Newton's constant. When considering compact objects as the lens population, the point-source point-lens (PSPL) model \citep{wambsganss2006gravitational}, where both the lens and the stellar source are treated as points, is a sufficient approximation as it characterizes the symmetric light curve model of microlensing events \citep{rektsini2024finding}. 

For essentially all photometric microlensing observations, the best measured parameter is the Einstein crossing time. The distribution of event timescales for a given population is determined by the lens mass function, spatial profile, and kinematics. However, the degeneracy of mass, distance, and velocity can be broken by additional microlensing observables. Astrometric microlensing can provide a direct measure of the angular Einstein radius, $\theta_{E}$, through the centroid motion in the source \citep{bramich2018predicted} and deviations in the symmetric light curve due to the Earth's motion determine the microlensing parallax, 
\begin{equation}
\pi_{E} = \frac{\pi_{\text{rel}}}{\theta_{E}}\,,
\end{equation}
where $\pi_{\text{rel}} = 1 \text{AU} (\frac{1}{D_{L}} - \frac{1}{D_{S}})$ is the relative lens-source parallax \citep{gould1992extending, alcock1995first}. The distributions of $t_{E}$, $\theta_{E}$ and $\pi_{E}$ are all sensitive to differences in the underlying lens population.

Fully realizing the constraining power of the upcoming microlensing surveys requires a dedicated modeling effort that differentiates physically distinct lens populations. The dark matter discovery potential is tempered by the challenge, already present with baryonic objects alone \citep{sajadian2019predictions, sajadian2023detecting, perkins2024disentangling}, of disentangling those populations. 

In this paper, we compare population-level microlensing observables for PBHs and DBHs. We determine how the population parameters of dark compact objects are imprinted on the statistics of the microlensing observables. In Section \ref{sec:Dark matter}, we provide additional background and details on the distinct formation channels for primordial black holes and dissipative dark matter objects. Section \ref{sec:micropop} describes the model for each potential dark lens population in the Milky Way. In Section \ref{sec:results}, we report the distribution of microlensing observables for each population. Section \ref{sec:discussion} discusses the broader context for these results, including the expected number of dark lenses, the overlap with microlensing from baryonic lenses, and complementarity with other gravitational probes. We conclude in Section \ref{sec:conclusion}.

\section{Dark Matter Models and Signatures}
\label{sec:Dark matter}
Dark matter, which constitutes nearly $85\%$ of all matter \citep{cirelli2024dark}, is required to explain a wide range of phenomena, from the rotation curves of spiral galaxies \citep{rubin1983dark} to the anisotropies of the cosmic microwave background \citep{aghanim2020planck} and the large-scale structure of the Universe \citep{blumenthal1984formation}. These observations tell us that dark matter is non-baryonic, but its particle nature remains a pressing question. A myriad of experimental detection efforts have been developed to search for dark matter candidates \citep{bertone2018new}, especially weakly interacting massive particles (WIMPs) \citep{akerib2020projected}. These searches continue to narrow the parameter space and constrain DM particle properties, but no detection of non-gravitational interactions between dark and Standard Model matter has been confirmed. In parallel, the detection of gravitational waves from merging compact objects by the LIGO-Virgo-KAGRA (LVK) Collaboration \cite{abbott2016observation} provides a new means to probe the microphysics of dark matter using gravity alone. The ever-growing catalog of gravitational wave events \citep{callister2024observed, abac2025gwtc} and the ability of current detectors to search for sub-solar mass ultra-compact objects has reignited interest in black holes of dark matter origin. 

PBHs have historically been the benchmark candidate for dark compact objects, and models that produce them have been extensively studied (see, for example, reviews including \citep{carr2020primordial, carr2022primordial, green:2026}). The potential PBH mass range is wide and largely unconstrained, spanning many orders of magnitude. The PBH mass, to first order, follows
\begin{equation}
    M_{\text{PBH}} \sim \frac{c^{3}t}{G}\,,
\end{equation}
\citep{carr1974black} where $c$ is the speed of light, $G$ is the gravitational constant and $t$ is the time of formation. Different inflationary models are required to generate different mass ranges for any PBH population, although final masses also depend on accretion \citep{carr2022primordial}. A number of papers have considered PBHs as a contributing source for some of LVK's black holes, in a variety of mass ranges \citep{Bird_2016,Sasaki:2016}. For simplicity, most works consider a delta-function or narrow Gaussian for the mass function of PBHs, with constraints on other distributions inferrable from there \citep{Bellomo:2018}.

Black holes of dissipative origin, DBHs, are a more recent topic of study and the cosmology, chemistry, and structure-formation dynamics possible in the wide class of scenarios that can produce them has only begun to be explored \citep{kaplan2010atomic, fan2013double, cyr2013cosmology, foot2013galactic, ghalsasi2018exploring, buckley2018collapsed, ryan2022molecular, gurian2022molecular, ryan2022molecularIII,Gurian:2022nbx,roy2023simulating, roy2025aggressively,schon2025shh}. The mass spectrum of these black holes will be bounded by a Chandrasekhar relationship like that of the Standard Model. In terms of the heaviest relevant fermion mass, $m_X$, that is 
\begin{equation}
    M^{\text{Dark}}_{\text{Chand.}} \propto 1.4 M_{\odot} \biggl ( \frac{m_{p}}{m_{X}} \biggr)^{2}
\end{equation}
where there is a constant depending on the number of degrees of freedom. However, the mass range of any dark black holes that do form above this limit is dependent on the detailed chemistry of the dark matter particles. Given the similarity with baryonic compact object formation, natural distributions to consider are those similar to stellar initial mass functions, especially if one supposes the dark sector is rather simpler than the baryonic sector (e.g., fewer cooling channels, no analog of nuclear physics). 

In addition to distinct mass functions, DBH and PBH populations are expected to exhibit different spatial and velocity distributions according to their underlying microphysics. PBHs, as collisionless dark matter, trace the Milky Way dark matter halo \citep{pruett2022primordial}, well-described by the Navarro-Frenk-White profile \citep{navarro1997universal}, with velocities derivable from the Eddington-like inversion method \citep{lacroix2018anatomy}. On the other hand, DBHs likely trace a different spatial distribution, according to dark matter regions where sufficient cooling could take place. Some dark compact objects could have formed very early in the core of our galaxy, or could have been transferred in through the Milky Way's merger history. The DBH population would likely be primarily distributed in the stellar bulge, with typical velocities similar to that of the baryonic population in this environment.

While neither PBH nor DBH scenarios are constrained to produce only light black holes, the subsolar-mass region offers a compelling target since baryonic stellar processes cannot produce black holes at such low masses. Motivated by this clear discovery potential, the LVK Collaboration and independent groups have conducted dedicated searches for compact binaries with masses between $0.01-1$ $M_{\odot}$ \citep{abbott2018search,abbott2019search,LIGOScientific:2021job, nitz2021search,LVK:2022ydq,nitz2022broad,kacanja2026search}. No statistically significant candidates have so far been found, but future improvements in detector sensitivity will enhance LVK's ability to detect mergers containing light objects farther out into the universe.

In addition, the $\mathcal{O}(1\,M_{\odot})$ mass range is interesting because heavier objects are easier to detect in mergers and because there are a number of light objects observed whose exact nature is unconfirmed. In the absence of associated electromagnetic signals or observed tidal deformability, gravitationally detected objects are often given a physical label according to their mass. Several gravitational wave events (GW191219, GW190425, GW200115, GW190917, GW190814, GW200210, and GW230529, \citep{abbott2020gw190425, mandel2021gw200115, sanger2024tests}) have at least one mass component below the expected lowest black hole mass from stellar processes (~3 $M_{\odot}$). These objects are therefore labeled as neutron stars (NSs) \citep{zhu2022population, abbott2019gwtc, abbott2023gwtc}, even though the data cannot actually distinguish them from black holes. Several works have explored the possibility that some of those events are indeed black holes from novel populations \citep{singh2021gravitational,clesse:2020ghq,khadkikar:2025}. 

Microlensing, similarly, cannot determine the exact nature of a lens based solely on the duration of the photometric light curve, whose shape has a degenerate dependence on mass, distance, and kinematics. This degeneracy is broken when additional microlensing observables, the angular Einstein radius from astrometric followup and the microlensing parallax, can also be obtained. Through a combination of photometric, astrometric, and parallax measurements, and the absence of detected lens flux, a dark lens candidate measuring $6$ $M_{\odot}$ (OGLE-2011-BLG-0462/MOA-2011-BLG-191) has been classified as the first isolated black hole in the Milky Way \citep{lam2022isolated, mroz2022systematic, sahu2022isolated, lam2023reanalysis}. More recently, several microlensing events detected with OGLE-III and followed up with astrometric measurements from Gaia have been reported as dark lens candidates occupying the subsolar mass range \citep{mroz2021measuring, kruszynska2024dark}. Since the nature of these lens candidates remains uncertain, the classification ambiguity leaves open the possibility that some events may arise from novel populations.

\section{Dark Lens Models}
\label{sec:micropop}
To explore how the intrinsic population properties map onto microlensing observables, we consider a single line of sight and draw lens masses, line-of-sight distances, and transverse velocities from their corresponding probability distribution functions: $P(m)$, $P(D_{L})$ and $P(D_{S})$, and $P(v_{\perp})$. Using the standard microlensing relations, we calculate the Einstein crossing time $t_{E}$, the angular Einstein radius $\theta_{E}$, and the microlensing parallax $\pi_{E}$ for each sampled microlensing configuration. This approach focuses on population-level statistics rather than individual event light curves and does not account for specific survey selection effects. It demonstrates how the different population parameters, motivated by unique lens population formation channels, are imprinted on the microlensing parameter space. Given the potential for definitive evidence for a new population from gravitational waves, we consider PBH and DBH populations with masses $0.1 \leq M \leq 1$ $M_{\odot}$.

\subsection{Milky Way Model}
We consider a simplified Milky Way model consisting of a dark matter halo and a central bulge, and we assume that all lenses are sampled along a one-dimensional line-of-sight, $(\ell, b) = (0,0)$, towards the Galactic Center. As all lens distances are measured from the observer, while the respective bulge and halo spatial profiles are defined in terms of the Galactocentric radius, we must consistently transform between coordinate systems. For a given lens distance $D_{L}$, the corresponding Galactocentric radius is given by $r(D_{L}) = |R_{0} - D_{L}|$, where $R_{0} = 8.3$ kpc is the distance to the Galactic Center from the Sun. For each simulated lens, a static source distance $D_S$ is drawn, sampled uniformly between $7-9$ kpc and satisfying $D_S > D_L$. 

The spatial distributions of lenses are modeled by analytic density profiles corresponding to specific regions of the Milky Way. The dark matter halo is modeled using the Navarro-Frenk-White (NFW) density profile 
\begin{equation}
    \rho_{\text{NFW}}(r) = \frac{\rho_{s}}{(1 + \frac{r}{r_{s}}) (1 + (\frac{r}{r_{s}}))^{2}}
\end{equation}
where $\rho_s = 0.0093$ $M_{\odot} \,  \text{pc}^{-3}$ is the scale density and $r_s = 18.6$ $\text{kpc}$ is the scale radius \citep{pruett2022primordial}. The Galactic bulge is modeled using the spherically symmetric de Vaucouleurs surface density profile given by
\begin{equation}
    \Sigma_{br}(r) = \Sigma_{b} \exp \biggl [ -k \biggl( \bigl ( \frac{r}{a_{b}} \bigr )^{1/4} - 1\biggr )\biggr]
\end{equation}
where $k = 7.6695$ and $\Sigma_{b} = 0.595 \times 10^9 M_{\odot} / \text{kpc}^{2}$ is the surface mass density at the  half-mass radius $a_{b} = 1.764$ $\text{kpc}$ fitted from the rotational velocities of stars and gas in the galaxy \citep{kumar2025exploring}.
Each spatial profile is utilized to construct a normalized probability density function $P(D_{L})$, from which lens distances are drawn via inverse transform sampling. 

Each lens population is assigned a Maxwellian transverse velocity distribution following 
\begin{equation}
    P(v_{\perp}) = \sqrt{\frac{2}{\pi}} \frac{v_\perp^2}{\sigma^{3}} \exp \biggl [ -\frac{v_\perp^2}{2\sigma^2}\biggr ]
\end{equation}
in accordance to average lens population kinematics. We truncate speeds at $v_{\text{esc}} = 550$ $\text{km/s}$ as this is escape velocity. Although $v_{\perp}$ is in reality dependent on the phase-space distribution of the Galactic potential, we will simplify and adopt a Maxwellian distribution with characteristic velocities separated by radial bins, as described in more detail below. Only the transverse velocity is sampled as we neglect source and observer's motion. This permits a faster sampling of lens population velocities.  

\subsection{Dark Compact Object Lenses}
We model two distinct populations of compact objects, PBHs and DBHs, each assumed to make up a fraction of dark matter in the Milky Way. Motivated by the classification ambiguity and discovery potential of gravitational probes (discussed at the end of Section \ref{sec:Dark matter}), we consider both novel populations to be in mass range $0.1-1$ $M_{\odot}$. 

For PBHs, we assume a truncated Gaussian distribution as is typical in the literature, centered at $\bar M_{\text{PBH}} = 0.5$ $M_{\odot}$, with $\sigma_{\text{PBH}} = 0.1$ $M_{\odot}$ with $M \in [0.1,1]\, M_{\odot}$. The population of PBHs is assumed to be a perfect (unbiased) tracer of the dark matter halo, with higher density near the Galactic Center. Therefore, the PBH distance distribution is sampled over the range $1.5 \leq D_{L} \leq 8.29$ kpc using the line-of-sight NFW profile. This population inherits the dark matter halo velocities which are approximated as a truncated Maxwellian with typical transverse speeds of $120$ $\text{km/s}$ for PBHs within $ r < 0.2$ $\text{kpc}$, $180$ $\text{km/s}$ for PBHs within $0.2 \leq r < 1.0$ $\text{kpc}$, $220$ $\text{km/s}$ for PBHs within $1.0 \leq r < 2$ $\text{kpc}$, and $250$ $\text{km/s}$ for PBHs at radii $r \geq 2.0$ $\text{kpc}$.

\begin{figure}[t]
    \centering
    \includegraphics[width=0.95\linewidth]{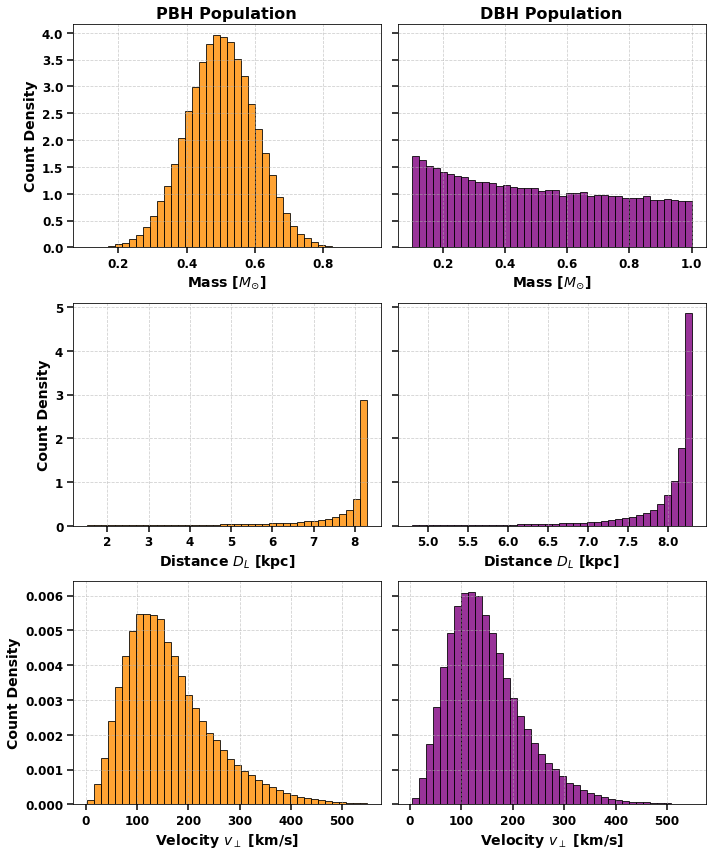}
    \caption{The sampled parameter distributions for the PBH population and the DBH population, including mass (top), distance from the observer (center), and velocity (bottom). PBHs are sampled using a Gaussian distribution centered at $0.5$ $M_{\odot}$ tracing the Navarro-Frenk-White halo profile, while DBHs are sampled using a power-law distribution $P(M) \propto M^{-0.3}$ with distances following the de Vaucouleurs bulge profile. Both populations sample transverse velocities from radially binned Maxwellian distributions with characteristic velocities ranging from $120$ to $250$ $\text{km/s}$. As each population is sampled from a distinct spatial distribution, the distance-dependent velocity prescription yields different effective kinematic distributions.} 
    \label{fig:Sampled Populations}
\end{figure}

Since DBH formation resembles that of baryonic compact objects, we assume a power-law distribution 
\begin{equation}
    P(M_{\text{DBH}}) \propto M_{\text{DBH}}^{-\beta_{D}}\,.
\end{equation}
We assume a mass range of $0.1 \leq M \leq 1$ $M_{\odot}$ and take $\beta_{D} = 0.3$ so that the average mass of the DBH and PBH populations is the same, $0.5$ $M_{\odot}$. We assume that most DBHs were either produced very early in the core of the Milky Way or brought in through mergers. In either case, they will not track the dark matter halo but instead preferentially reside today within the stellar bulge which we take to have radius of $3.5$ $\text{kpc}$, following de Vaucouleurs profile. The DBHs therefore occupy distances between $4.8 \leq D_{L} \leq 8.29$ $\text{kpc}$, with the upper bound for consistency with the PBH distance distribution. The DBH transverse velocity distribution follows the distance-dependence prescription as the PBH population. Since the two populations are spatially distinct, their corresponding velocity distributions are also different. Figure \ref{fig:Sampled Populations} shows the mass, radial position, and velocity distributions for each population.

\section{Results}
\label{sec:results}
\begin{figure}[t]
    \centering
    \includegraphics[width=0.95\linewidth]{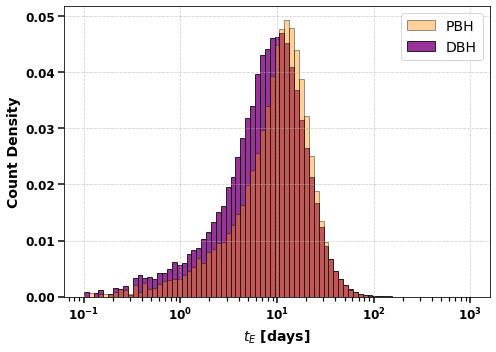}
    \caption{The Einstein crossing time distribution of primordial black holes (PBHs) and dark black holes (DBHs). While the populations share similar timescale averages near $20$ days, their distribution also reveals statistical differences in the spread and tails, indicating distinct underlying population properties.}
    \label{fig:DCOs Crossing Time}
\end{figure}

The Einstein crossing time is the most commonly measured microlensing observable as it is directly inferred from photometry alone. It depends on lens mass, lens-source separation, and kinematics, and even in the absence of astrometric follow-up or parallax measurements, may provide a statistical signature of the underlying population properties. Figure \ref{fig:DCOs Crossing Time} shows that both DBH and PBH populations defined above would produce events with characteristic timescales between $1-100$ days, with a peak near $15$ days. While the average duration is the same for both populations, DBHs exhibit a distinct shape and spread in the timescale distribution primarily towards the short timescale tail since PBHs trace the dark matter halo while DBHs are typically further away from the observer. 

\begin{figure*}
    \centering
    \includegraphics[width=0.95\linewidth]{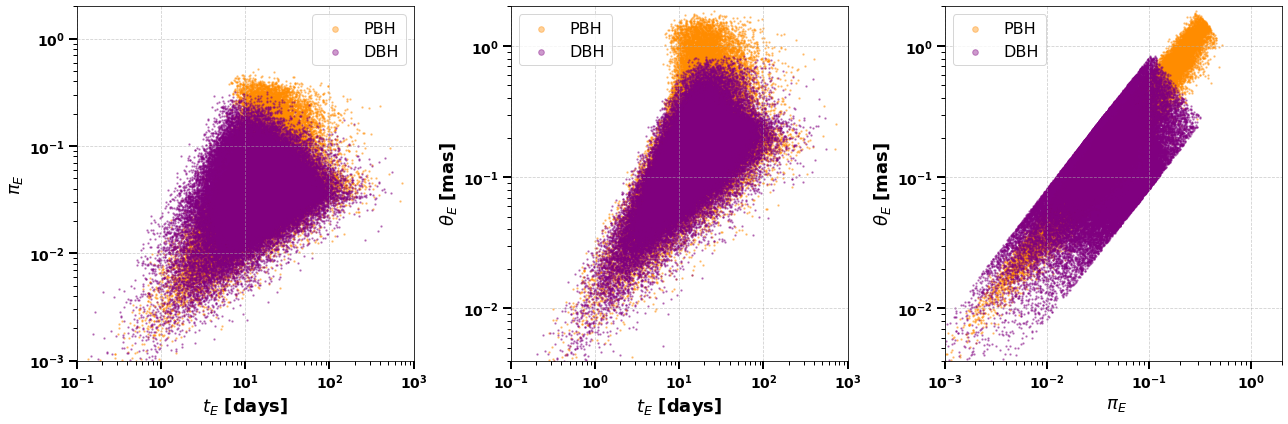}
    \caption{The joint parameter spaces for the Einstein crossing time $t_{E}$, parallax $\pi_{E}$, and angular Einstein radius $\theta_{E}$, for primordial black holes (PBHs, orange) and dark black holes (DBHs, purple). The primary difference between the two populations comes from their spatial distribution: PBHs trace the halo, while DBHs are clustered in the bulge.}
    \label{fig:Microlensing DCO Parameter Space}
\end{figure*}

The Einstein crossing time alone cannot be used to separately distinguish the mass, velocity or distance of the lens. Some of this degeneracy can be broken by jointly considering the larger set of observables, $t_{E}$, $\theta_{E}$, and $\pi_{E}$. Figure \ref{fig:Microlensing DCO Parameter Space} shows where the DBHs and PBHs sit in the joint parameter spaces $(t_{E}, \pi_{E})$, $(t_{E}, \theta_{E})$, and $(\pi_{E}, \theta_{E})$. The left panel shows that the PBH population, which extends closer to the observer while DBHs primarily populate the bulge, has objects with higher parallax at fixed crossing time. The middle panel similarly shows that the DBH population is limited to smaller angular Einstein compared to PBHs. The right panel shows how the spatial distribution affects $\theta_{E}$ and $\pi_{E}$. The microlensing parallax and the angular Einstein radius both increase with increasing lens-source separation, so the PBHs outside the bulge are cleanly separated.

\section{Discussion}
\label{sec:discussion}
\subsection{Abundance of Dark Black Holes in Near-future Surveys}
Of course, a novel lens population must be sufficiently abundant to produce observable events in upcoming surveys. For DBHs, this requires that the fraction of dark matter in DBHs in the Milky Way, $f^{\text{MW}}_{\text{DBH}}$, be high enough for the number density along the line-of-sight to produce detectable events. The number of isolated DBHs in the Milky Way is approximately given by 
\begin{equation}
    N_{\text{DBH}} = \frac{f^{\text{MW}}_{\text{DBH}}M_{\text{DM}}}{\langle m_{\text{DBH}}\rangle} (1-f_{\text{binary}})
\end{equation}
where $M_{\text{DM}} = 10^{12}$ $M_{\odot}$ is the total dark matter mass in the Milky Way, $\langle m_{\text{DBH}}\rangle$ is the average DBH mass, and $f_{\text{binary}}$ is the binary fraction. 

To motivate performing a more detailed forecast for upcoming surveys, we obtain an order-of-magnitude estimate for the number of detectable DBHs, based on prior results in the literature. First, we use the recent detailed PopSyCLE analysis \citep{lam2020popsycle} to estimate an event rate for Galactic bulge lines of sight. We use the white dwarfs from \cite{lam2020popsycle} as a reference population, since their mass scale and spatial distribution is comparable to that of our $\langle m_{\text{DBH}}\rangle = 0.5$ $M_{\odot}$ DBHs. For the Early Warning System (EWS) OGLE survey, \cite{lam2020popsycle} found roughly one lensing event for every $10^6$ white dwarfs. This suggests that for $f_{\text{binary}}=1/3$ and $f^{\text{MW}}_{\text{DBH}} = 10^{-5}$ (well below current gravitational wave constraints \citep{abbott2022search,singh2021gravitational}), a galactic population of about $1.5 \times 10^7$ isolated DBHs would contribute $\mathcal{O}(10)$ events with well-constrained masses in that survey. The Roman Space Telescope will probe the Galactic bulge with high-cadence photometry improving the detection rate of microlensing events with mass measurements by 1-2 orders of magnitude \citep{lam2020popsycle}. In addition to photometry, the number of detected $1$ $M_{\odot}$ events through purely astrometric microlensing has an upper limit of around $10^3 f_{\text{DM}}$ \citep{fardeen2024astrometric}. These estimates find the potential for a significant population of detected DBHs, indicating it would be worthwhile to perform a more precise forecasting analysis for DBH populations and next-generation surveys.  

While we so far only considered isolated objects as lenses, the total compact object population includes a large number of binaries. The binary fraction is a fundamental population parameter determined by formation physics and dynamical evolution. Similarly to stellar-origin binaries, the fraction of dark-origin binaries depends on the spectrum of primordial fluctuations, any chemistry of the dark matter, and standard gravitational kinematics \citep{shandera2018gravitational,raidal2019formation}. For example, in the atomic dark matter scenario, which has only hydrogenic chemistry, formation of objects and binaries mimics Pop III star formation closely \citep{shandera2018gravitational}. Recent analyses \citep{abrams2025assessing,bozza2025historic} indicate that about half of all microlensing events will contain a binary system, either for the lens or the source. The non-merging fraction of the DBH population may therefore still be accessible with microlensing observations, although an analysis of binary parameters and signal is needed. The merging fraction can be directly probed by gravitational wave observatories.

\subsection{Complementary Gravitational Probes}

\begin{figure}[ht]
    \centering
    \includegraphics[width=0.95\linewidth]{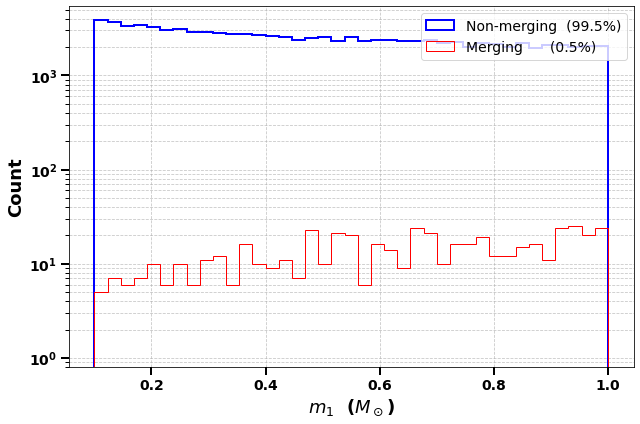}
    \caption{An example distribution of mass of the primary (heavier) black hole in dark black hole binaries that merge within the Hubble time (red) compared to the distribution of all black holes (blue). The plot uses the atomic dark matter scenario, where DBH formation approximately follows the dynamics of Pop III star formation studied in \citep{shandera2018gravitational} but with the binary parameter distributions adjusted to the mass function considered in this paper. A sample of 100,000 dark black holes was used to generate the comparison. \label{fig:Merging Fraction of DBH Binaries}}
\end{figure}

This complementary nature of microlensing and gravitational waves plays a crucial role in studies of compact object populations. Gravitational wave observatories are biased towards more massive, tightly bound binaries that merge within the age of the universe. The formation mechanism of dark matter-origin compact objects will also determine the orbital parameter distributions and the merger rate. Assuming binary parameters as given in \citep{shandera2018gravitational} for atomic dark matter, the DBH mass function used in this paper would yield a low merging fraction as shown in Figure \ref{fig:Merging Fraction of DBH Binaries}. While this fraction is low, it generates a non-negligible number of events for $f_{\text{DBH}} \gtrsim 10^{-5}$ and the current sensitive volume of the LVK detectors. Indeed, previous work showed that interpreting the stellar-mass components in GW190425 and GW190814 as DBHs in the context of atomic dark matter gives $f_{\text{DBH}}$ between $10^{-5}$ and $10^{-3}$. This interpretation also bounds the mass of the heavier dark fermion to be heavier than the Standard Model proton, showing how gravitational detections can constrain microphysics \citep{singh2021gravitational}. 

Since microlensing and gravitational wave techniques probe different subpopulations across different environments, constraints on the fraction of dark matter in compact objects must be interpreted keeping that context in mind. Microlensing surveys probe the abundance of DBHs within the specific environment of the Milky Way, which will have a different fraction of dark matter in dark compact objects than the extragalactic volume probed by gravitational wave detectors will. The fraction of collapsed dark matter in any galaxy, $f^{\rm local}_{\text{DBH}}$, depends on the local dark chemical history and so will differ with galaxy morphology and age. Dissipative dark compact objects are biased tracers of the underlying matter density, just as baryonic objects are. Microlensing and gravitational wave constraints are related in a model-dependent way for DBHs, while more straightforwardly related for unbiased tracers like PBHs. 

\subsection{Baryonic Lens Overlap}
\begin{figure}[t]
    \centering
    \includegraphics[width=0.95\linewidth]{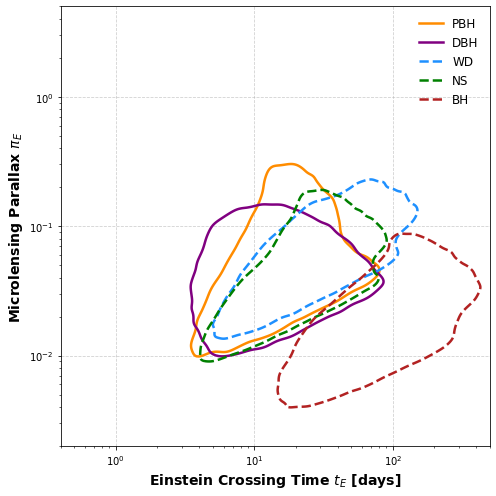}
    \caption{The $90\%$ kernel density estimate contour for the sampled dark compact object populations (primordial black holes and dark black holes) and baryonic compact object populations (white dwarfs, neutron stars, black holes) in the joint $(t_{E}, \pi_{E})$ parameter space. The low-mass stellar remnants, white dwarfs and neutron stars, strongly overlap with the dark compact objects. This overlap further motivates precise forward-modeling to disentangle populations.}
    \label{fig:Full Microlensing Parameter Space}
\end{figure}

While DBHs and PBHs may occupy distinct regions of the parameter space, they may overlap with baryonic objects for any given observable. Stellar remnants, including white dwarfs (WDs), neutron stars (NSs), and stellar-origin black holes (SOBHs) provide the baseline microlensing populations against which the dark compact object population must be compared to. We next estimate this overlap. Following the population prescription of baryonic compact objects from \citep{lam2020popsycle}, we show in Figure \ref{fig:Full Microlensing Parameter Space} that there are distinct regions where stellar remnants do not occupy which DCOs (of which we've considered DBHs and PBHs) do.

In the Milky Way, stellar remnants are expected to maintain the same spatial distributions as their progenitor stars, mainly in the stellar bulge and disk. As in \citep{lam2020popsycle}, we consider microlensing events along the line of sight towards the Galactic Center where the stellar remnant population considered lives in the bulge and the disk. In addition to the same bulge profile used for DBHs, the disk is modeled with an exponential profile,
\begin{equation}
    \rho_{\text{disk}}(r) \propto \exp \left[ - \frac{(r - R_{0})}{R_d}\right]\ ,
\end{equation}
with the scale length $R_d = 2.5$ $\text{kpc}$, and distance to the Sun, $R_{0} = 8.3$ $\text{kpc}$. To model the baryonic stellar remnants alongside the dark compact objects, we assume that $80\%$ of each baryonic lens population resides in the bulge. The radial ranges of the bulge and disk partially overlap to produce a smooth transition between the two components and avoid artificial hard boundaries when sampling lens distances. Bulge velocities are typically higher than disk velocities so the fraction of stellar remnants in the bulge have higher average speeds. In addition, neutron stars and stellar-origin black holes receive natal kicks during formation, so we increase the mean transverse velocity for these populations. 

The stellar remnant mass function provided by stellar evolution models provides a baseline for the expected compact object mass distribution, with white dwarfs in the $0.2-1.4$ $M_{\odot}$ range \citep{gould2000measuring}, neutron stars in the $1-2$ $M_{\odot}$ \citep{ozel2016masses} range, and black holes in the $> 4$ $M_{\odot}$ range \citep{farr2011mass}. The corresponding velocity distributions are therefore determined from the dynamics of the stellar environment and natal kicks during formation \citep{popov2025natalkickscompactobjects}. 

We draw white dwarf masses from a truncated Gaussian with average mass $\bar M_{\text{WD}} = 0.6$ $M_{\odot}$ with standard deviation $\sigma_{\text{WD}} = 0.1$ $M_{\odot}$ with $0.2 \leq M \leq 1.4$ $M_{\odot}$. The bulge WDs are given Maxwellian transverse velocity distribution with mean transverse velocities of $140$ $\text{km/s}$ while the disk WDs have mean transverse velocities of $70$ $\text{km/s}$ \citep{bland2016galaxy}. We take the neutron star mass function to follow a two-component truncated Gaussian with a narrow peak at $1.4$ $M_{\odot}$ and a broader peak at $1.8$ $M_{\odot}$ with $1.2 \leq M \leq 2.3 $ $M_{\odot}$ \citep{you2024birth}. However, NSs receive natal kicks during formation which motivates Maxwellian distributions of transverse velocities to be higher, $\bar v_{\perp,\text{bulge}} = 300$ $\text{km/s}$ and $\bar v_{\perp,\text{disk}} = 250$ $\text{km/s}$ \citep{disberg2025kick}. For the stellar-origin black hole population, we follow \citep{lam2020popsycle} and adopt a truncated power-law mass function $P(M) \propto M^{-\alpha}$ where $\alpha = 1.6$ between $5 \leq M \leq 16$. Black holes typically receive smaller natal kicks than NSs so the Maxwellian distribution has characteristic speeds of $200$ $\text{km/s}$ for the bulge and $120$ $\text{km/s}$ for the disk. 

As most microlensing event classification depends on model-dependent inferences that assume only baryonic objects, the stellar remnant population stands as the baseline to compare to microlensing event anomalies to. These results show that WDs and NSs will share regions of the parameter space with DCOs, further motivating proper modeling of all objects occupying this low-mass range. The heavier population consisting of SOBHs is largely distinguishable due to its long-duration timescales and low microlensing parallax. With microlensing event detections expected to increase by orders of magnitude, a subset of those events may exhibit characteristics of observables that baryonic objects cannot explain reasonably well. These population-level analyses are crucial to constrain populations and the nature of dark matter. 

While DCOs and baryonic stellar-origin compact objects may occupy distinct regions in the microlensing parameter space, there exist other non-luminous astrophysical populations that could overlap with the subsolar DCOs, especially the free-floating planets (FFPs) and isolated brown dwarfs. Since the crossing time $t_{E} \propto \sqrt{M}$, these low-mass populations also generate short microlensing timescale distributions and may overlap with DCOs. Typically, microlensing surveys interpret short timescale events as FFP or brown dwarf candidates, introducing classification ambiguity and creating background contamination in classification studies of DCOs in microlensing data.

While previous studies \citep{derocco2024revealing} have contrasted the statistical signatures of PBHs with FFPs, these FFPs also overlap the DBH population in the subsolar mass regime, albeit with different mass ranges and spatial and velocity distributions. This implies that while there may be overlap in the short timescale distribution of microlensing events, the joint microlensing observable space  $(t_{E}, \theta_{E}, \pi_{E})$ could also help in distinguishing between DBHs and other subsolar populations. FFPs and brown dwarfs have distinct spatial and velocity priors that will aid in statistically distinguishing them from DBHs. With the expected increasing statistics of upcoming microlensing catalogs, this classification ambiguity prompts proper hierarchical inference to statistically disentangle overlapping populations. 

\section{Conclusion}
\label{sec:conclusion}

In this work, we computed microlensing observables $(t_{E}, \theta_{E}, \pi_{E})$ for two different, physically motivated DCO populations, primordial black holes (PBHs) and dark black holes (DBHs) formed from dissipative dark matter. These have distinct mass functions, spatial profiles, and distance-dependent velocities distributions dependent on their formation channels. When the two populations overlap in mass, the primary differences in the resulting microlensing parameter space come from the distinct spatial distributions. This is most clearly seen in the $(\pi_{E}, \theta_{E})$ microlensing parameter space, as shown in Figure \ref{fig:Microlensing DCO Parameter Space}. Rescaling current microlensing simulation results suggest that even a subdominant fraction of dark matter in DBHs, $f^{\text{MW}}_{\text{DBH}} = 10^{-5}$, below current gravitational wave limits, could yield $\mathcal{O}(10)$ microlensing events. This suggests that astrometric measurements, in addition to photometric observations, will play a major role in distinguishing dark compact object populations. These results also show that realizing the full potential of microlensing as a probe of dark matter, and especially interpreting any hints of an anomalous population towards the Galactic bulge \citep{perkins2025hints}, requires modeling populations beyond PBHs.

While the origin of any single black hole, detected by lensing or in a merger, cannot be determined, a wide range of other observations can test the dissipative dark matter origin story. Dissipative models have rich phenomenological consequences beyond the possible production of new compact objects, especially for structure formation on sub-galactic scales \citep{brito2022snowmass2021}. Radiative cooling in a dissipative dark sector would modify halo structure in a model-dependent way, which has only begun to be explored with dedicated simulations \citep{roy2023simulating, roy2025aggressively, rose2024introducing, schon2025shh, gemmell2024dissipative, guerra2026probingatomicdarkmatter}. While detailed predictions of dark compact object populations from first-principles remain challenging, the order-of-magnitude improvement in observational constraints, or any detections, will be a key piece of data driving further study. 

Acknowledgments: We are grateful for illuminating conversations with Joshua Black, Donghui Jeong, and Jessica Lu. This work was supported by NASA under ATP 80NSSC22K0819.

\newpage 
\clearpage
\bibliographystyle{aasjournal}
\bibliography{references}
\end{document}